\documentclass[conference]{IEEEtran}

\usepackage{acronym}
\usepackage{algorithm}
\usepackage{algpseudocode}
\usepackage{amssymb}
\usepackage{colortbl}
\usepackage{graphicx}
\usepackage{mathtools}
\usepackage{multicol}
\usepackage{multirow}
\usepackage{orcidlink}
\usepackage{pseudocode}
\usepackage{rotating}
\usepackage{soul}
\usepackage{tabularray}

\usepackage[dvipsnames]{xcolor}

\acrodef{AI}{artificial intelligence}
\acrodef{API}{application programming interface}
\acrodef{APK}{android package kit}
\acrodef{BERT}{bidirectional encoder representation for transformers}
\acrodef{BFS}{breadth-first search}
\acrodef{CFG}{control flow graph}
\acrodef{CNN}{convolutional neural network}
\acrodef{CPU}{central processing unit}
\acrodef{DAGAE}{distribution-regularized adversarial graph autoencoder}
\acrodef{DeDe}{Delphi Decompiler}
\acrodef{DEX}{Dalvik executable format}
\acrodef{DFS}{depth-first search}
\acrodef{DLL}{dynamic link library}
\acrodef{DNN}{deep neural network}
\acrodef{ELF}{executable and linkable format}
\acrodef{FCG}{function call graph}
\acrodef{FPR}{false positive rate}
\acrodef{GNN}{graph neural network}
\acrodef{JSON}{JavaScript Object Notation}
\acrodef{GAN}{generative adversarial network}
\acrodef{GIN}{graph isomorphism network}
\acrodef{GPT}{generative pre-trained transformer}
\acrodef{GPU}{graphics processing unit}
\acrodef{HGBC}{histogram gradient boosting classifier}
\acrodef{HPC}{high performance computing}
\acrodef{IoT}{Internet of things}
\acrodef{KNN}{K-nearest neighbors}
\acrodef{LIME}{Local Interpretable Model-agnostic Explanations \cite{ribeiro_why_2016}}
\acrodef{LLM}{large language model}
\acrodef{LSTM}{long short-term memory}
\acrodef{MalGAN}{Malware Generative Adversarial Network \cite{hu_generating_2017}}
\acrodef{ML}{machine learning}
\acrodef{NLP}{natural language processing}
\acrodef{OS}{operating system}
\acrodef{OSINT}{open source intelligence}
\acrodef{PE}{portable executable}
\acrodef{PGD}{projected gradient descent}
\acrodef{RNN}{recurrent neural network}
\acrodef{RoBERTa}{Robustly Optimized BERT Pretraining Approach \cite{liu_roberta_2019}}
\acrodef{SHAP}{SHapley Additive exPlanations \cite{lundberg_unified_2017}}
\acrodef{SVM}{support vector machine}
\acrodef{VM}{virtual machine}
\acrodef{XAI}{eXplainable AI}

 \makeatother

\begin{document}
       

\title{Explainability-Guided Adversarial Attacks on Transformer-Based Malware Detectors using Control Flow Graphs}

\author{\IEEEauthorblockN{Andrew Wheeler}
\IEEEauthorblockA{\textit{Department of Computer Science} \\
\textit{Tennessee Tech University}\\
Cookeville, USA \\
amwheeler43@tntech.edu}
\and
\IEEEauthorblockN{Kshitiz Aryal}
\IEEEauthorblockA{\textit{School of Interdisciplinary Informatics} \\
\textit{University of Nebraska Omaha} \\
Omaha, USA \\
karyal@nebraska.edu}
\and
\IEEEauthorblockN{Maanak Gupta}
\IEEEauthorblockA{\textit{Department of Computer Science} \\
\textit{Tennessee Tech University}\\
Cookeville, USA \\
mgupta@tntech.edu}}

\maketitle

\begin{abstract}
Transformer-based malware detection systems operating on graph modalities such as \acp{CFG} achieve strong performance by modeling structural relationships in program behavior. However, their robustness to adversarial evasion attacks remains underexplored. This paper examines the vulnerability of a RoBERTa-based malware detector that linearizes \acp{CFG} into sequences of function calls, a design choice that enables transformer modeling but may introduce token-level sensitivities and ordering artifacts exploitable by adversaries. By evaluating evasion strategies within this graph-to-sequence framework, we provide insight into the practical robustness of transformer-based malware detectors beyond aggregate detection accuracy.

This paper proposes a white-box adversarial evasion attack that leverages explainability mechanisms to identify and perturb most influential graph components. Using token- and word-level attributions derived from integrated gradients, the attack iteratively replaces positively attributed function calls with synthetic external imports, producing adversarial CFG representations without altering overall program structure. 
Experimental evaluation on small- and large-scale Windows Portable Executable (PE) datasets demonstrates that the proposed method can reliably induce misclassification, even against models trained to high accuracy. Our results highlight that explainability tools, while valuable for interpretability, can also expose critical attack surfaces in transformer-based malware detectors. 


\end{abstract}


\IEEEpeerreviewmaketitle

\section{Introduction and Motivation}

\IEEEPARstart{T}{he} increasing volume and sophistication of malware has motivated the adoption of \ac{ML}-based techniques for automated malware detection. Traditional antivirus systems primarily rely on signature-based detection, which identifies malware by matching known byte patterns or hashes associated with previously analyzed samples. While effective for known threats, this approach struggles to detect new or modified malware variants because it requires signatures to be created and distributed before detection is possible. \Ac{ML}-based malware detection addresses this limitation by learning statistical and behavioral patterns that distinguish malicious software from benign programs. Consequently, \ac{ML}-driven detection systems have become an increasingly important component of modern cybersecurity defenses, particularly for identifying zero-day threats and evolving malware families in dynamic environments.


A major factor when training a malware detection system using \ac{ML} is determining how the binary executable file will be represented, a choice commonly referred to as selecting the \textit{modality}. This modality defines the form and structure of the features that will be extracted from an executable, and has important implications for the overall design of the detection pipeline. The selected representation not only dictates the feature extraction framework, but also influences the choice of model architecture itself, since different modalities are best handled by different types of models (\emph{e.g.} convolutional neural networks and image processing). 
There are modalities, such as raw byte sequences of binary executable, that needlessly consume superfluous data that is not relevant to the classification tasks. This is also the case with images as well, which are often derived from the raw byte sequences of executable files. Control Flow Graphs (\Acp{CFG}) solve this issue by reducing the amount of information that is extracted from the executable file while still maintaining structural characteristics.
Among the various modalities available, \acp{CFG} have emerged as a particularly promising representation. \acp{CFG} model the executable as a directed graph in which nodes correspond to basic blocks of code and edges represent possible execution paths, thus preserving both the structural layout of the program and important aspects of its runtime behavior. This ability to reflect program control flow makes \acp{CFG} especially useful for distinguishing benign from malicious behavior in executable code, as can be seen in \cite{muzaffar_android_2023} and \cite{bobrovnikova_technique_2022}.

One of the recent advancement in ML, the transformer architecture is a powerful model intended for language-based tasks, such as \ac{NLP} and sentiment analysis \cite{vaswani_attention_2023}. However, due to model flexibility, it can be adapted for other tasks it originally wasn't designed for, such as malware detection. In malware analysis, \acp{CFG} naturally contain textual information in the form of function names and external imports, allowing them to be linearized into text sequences derived from program call graphs and processed as a text classification task by transformer-based models.

Mathematically, we can represent a binary program as a graph $G$ with a set of vertices $V(G)$ corresponding to executable components of the program and a set of directed edges $\overrightarrow{E}(G)$ representing some relationship between elements of $V(G)$. A single vertex $v \in V(G)$ typically represents some executable portion of the program. This portion can be as small and atomic as a single assembly instruction, or it can be abstracted to higher levels, such as functions in the original source code. For every element $v \in V(G)$ there is at least one $e \in \overrightarrow{E}(G)$ for which it is an endpoint for. The directed edge connecting two program components typically represents a point at which the flow of execution is handed off from one component to the next. Organizing an executable program in this fashion allows for concise and simple representation of the program while still preserving the general control flow and form. Additionally, it allows for abstracting away from file format-specific structuring, such as if a program is formatted as a \ac{PE} for Windows or \ac{ELF} for Linux. 
In graph based ML malware detection systems, structural and semantic information derived from vertices and edges, such as node features, instruction sequences, or connectivity patterns, are used as inputs to a learning model \cite{ma_combination_2019, muzaffar_android_2023, bobrovnikova_technique_2022}. By leveraging both the structural relationships encoded in $\overrightarrow{E}(G)$ and the semantic information contained in $V(G)$, graph-based models can capture program behavior more effectively than flat feature representations, enabling improved generalization to previously unseen malware samples.


\textbf{Adversarial Attacks on Malware Detector:}
While ML-based malware detectors enable   zero-day malware detection, they open up a threat surface for adversarial evasion (AE) attacks. Such AE attacks are performed by introducing a carefully crafted adversarial perturbations to the test samples such that malware is identified as benign by detectors while it still preserves its malicious functionality.
Existing adversarial evasion attacks against malware detectors have largely focused on byte-based or feature-vector representations of binaries. In these approaches, adversaries typically manipulate raw byte sequences, inject benign byte patterns, or modify extracted features in order to evade detection while preserving program functionality~\cite{kolosnjaji2018adversarial,aryal2024survey,aryal2024explainability,aryal2025intra}. Such attacks have been widely studied in the context of deep learning models that operate directly on binary byte streams or static feature representations, demonstrating that small perturbations can significantly reduce detection accuracy while maintaining executable behavior. These techniques, however, are closely tied to the representation used by the detector and therefore do not directly translate to models that operate on structural program representations.


While graph based malware detection approaches capture richer semantic and behavioral information than byte-level representations, adversarial evasion attacks against them remain relatively limited. Existing work on graph-based attacks typically focuses on modifying graph structure or node attributes to influence model predictions, but these studies often target graph neural networks rather than transformer-based architectures \cite{ma_combination_2019, muzaffar_android_2023, bobrovnikova_technique_2022}. As a result, there remains a gap in understanding how adversarial evasion strategies can be applied to malware detectors that leverage transformer models over graph-derived representations. 

Our work addresses this gap by investigating adversarial evasion attacks against a transformer-based malware detector that operates on CFG-derived sequences, providing insight into the robustness of this increasingly adopted detection paradigm. We propose the use of \ac{XAI} techniques to provide the insight for attacks by identifying which components of an input contribute most strongly to a model’s classification decision. \Ac{XAI} informs us of which function calls within a \ac{CFG} contribute the most to an executable's malicious classification. By identifying and modifying these functions, we can evade a malware detection system through a tailor-made attack for the model.

The following are the key contributions of this paper:
\begin{itemize}
    \item We propose a novel process for generating adversarial malware samples via \ac{CFG} modification utilizing a multi-iteration attack.
    \item We utilize attributions generated by integrated gradients of a complex transformer model to inform generation of adversarial samples of precompiled Windows PE files against a white-box target.
    \item We launch an adversarial attack against a RoBERTa classification model capable of detecting malware with a high degree of accuracy.
    \item We analyze the real-world relevancy of the attack and propose methods by which it could be implemented in a practical setting.
\end{itemize}

The remainder of this paper is organized as follows. Section \ref{sec:PreviousWorks} presents the background and related work. Section \ref{sec:Methodology} describes the methodology. Section \ref{sec:Results} discusses the experimental results and analysis. Finally, Section \ref{sec:Conclusion} concludes the paper.

\section{Background and Related Works}\label{sec:PreviousWorks}

\begin{table*}[]
\centering
\resizebox{\textwidth}{!}{%
\begin{tabular}{c|c|c|c|l}
Paper &
  Model Architecture &
  Modality &
  Attack Context &
  \multicolumn{1}{c}{Contributions} \\ \hline
Raff et al. \cite{raff_malware_2017} &
  Neural Network &
  Raw Byte Sequences &
  N/A &
  Creation of advance malware classifier for large scale classification. \\
\rowcolor[HTML]{EFEFEF} 
Nicosia et al. \cite{nicosia_generative_2023} &
  MalGAN &
  CFG &
  Black-box &
  Modification of the MalGAN framework for control flow modalities. \\
\rowcolor[HTML]{FFFFFF} 
Yan et al. \cite{yan_litexgnn_2025} &
  GNN &
  CFG &
  N/A &
  Implements global and local explainability for CFGs. \\
\rowcolor[HTML]{EFEFEF} 
Shoukouhinejad et al. \cite{shokouhinejad_consistency_2025} &
  GNN &
  CFG &
  N/A &
  Utilization of advance explainability techniques for control flow modalities. \\
Soi et al. \cite{soi_enhancing_2024} &
  DNN &
  FCG &
  N/A &
  Augmentation of FCGs via inclusion of DEX information from APK files. \\
\rowcolor[HTML]{EFEFEF} 
Govea et al. \cite{govea_hybrid_2025} &
  GNN + CyberBert &
  \begin{tabular}[c]{@{}c@{}}Network Logs +\\ Traffic Flow\end{tabular} &
  N/A &
  Development of robust detector model using privacy-preserving federated learning. \\
Wheeler et al. \cite{wheeler_graphene_2026} &
  RoBERTa &
  FCG &
  N/A &
  Implementation of RoBERTa-based detector with high level of performance. \\
\rowcolor[HTML]{EFEFEF} 
Abusnaina et al. \cite{tagarelli_subgraph-based_2019} &
  \begin{tabular}[c]{@{}c@{}}CNN\\ DNN\end{tabular} &
  CFG &
  White-box &
  Adversarial attack on detector models built for IoT malware. \\
Jiang et al. \cite{jiang_shield_2026} &
  LLM &
  \multicolumn{1}{l|}{Program Source Code} &
  Black-box &
  Black-box attack on LLM-based vulnerability detectors.
  \\
\textbf{Our Work} &
  \textbf{RoBERTa} &
\textbf{CFG} &
  \textbf{White-box} &
  \textbf{Adversarial attack on Roberta-based graph modality malware detector.}
  \vspace{2mm}
\end{tabular}%
}
\caption{Summary of related works}
\label{tab:ReferencesOverview}
\end{table*}

Table \ref{tab:ReferencesOverview} provides an overview of related works for this paper. The individual works from the table are discussed in the sections below.

\subsection{Malware Detection Using ML}

The work of \cite{nicosia_generative_2023} highlights the role \acp{GAN} can play in creating adversarial malware samples using behavior based \acp{CFG}. The authors of this paper build on \ac{MalGAN}, a framework designed to generate adversarial malware samples to defeat \ac{ML}-based detector systems in a black-box context. They modify the existing \ac{MalGAN} codebase to use \ac{API} calls made to the \ac{OS} and use that as the feature vector. While this paper provides a graph-based extension to a robust adversarial system, the authors perform the attack on a standard neural network instead of a transformer system, and they utilize dynamic analysis through Cuckoo Sandbox instead of static analysis.

The \ac{RoBERTa} architecture is a transformer-based neural network designed for the task of \ac{NLP}. An important aspect of \ac{RoBERTa} is that is can be trained to perform text-based classification tasks with a high degree of accuracy. The model is able to accomplish this by tokenizing the input text sequences, embedding the tokens into a higher dimensional space, and ascribing contextual relationships between each of the tokens using multi-head self attention mechanisms. The contextual information is provided in the form of an attention mask, with each token having a unique attention masks that connects it to associated tokens in the original sequence. It is because of this mechanism for connecting relevant parts of a sequence that makes \ac{RoBERTa} such a powerful model, and by treating a given problem as a language-based problem, this architecture can be adapted to a large array of applications.

There is already an extensive body of work that evaluates the applicability of transformer-based malware detection models, as is described in the analysis paper of Kunwar et al. \cite{kunwar2025sok}. The architecture can be employed in a number of complex detection tasks, including malicious domain name detection, \ac{IoT} network intrusion safeguards, and Android malware classification. This flexibility coupled with the capacity for the model architecture to achieve a high level of performance makes it an incredibly powerful tool for classification tasks.

Graph-based transformers extend to domains outside of just malware detection, with federated models being designed for network intrusion detection. The work presented in \cite{govea_hybrid_2025} utilizes a multi-agent distributed detection system that combines a \ac{GNN}, a variation of \ac{BERT} known as CyberBERT, and a privacy-preserving federated learning framework. The authors provide a robust framework capable of strong performance in network intrusion detection. However, this work is focused on network intrusion detection and uses a multi-agent topology where explanations are provided for two models in tandem with each other.

Our previous work \cite{wheeler_graphene_2026} outlines the process and efficacy of using a \ac{RoBERTa} \ac{ML} system for the task of binary malware detection. However, this paper focuses solely on the performance of the model in conjunction with the effect traversal algorithms have on the ability for a model to accurately classify a sample. The robustness and ability of the model to withstand adversarial attacks is not evaluated, and no investigation into potential strategies for generating an adversarial control flow graph is made.

\subsection{Explainability of CFG Models}

The paper by Yan et al. \cite{yan_litexgnn_2025} describes a malware detector system that provides explanation-based analyses to augment detections. They utilize a \ac{GNN} as the detecting component of the system and perform inferencing to emphasize nodes and edges that contribute to a program being classified as malicious. Their approach is notable as it allows for both local and global explainability to better understand why the model ended up making a particular decision. However, even though this approach is effective in generating graph-based explanations, the authors do not extend the use of the explainability mechanism to implement an adversarial attack.

The work of \cite{shokouhinejad_consistency_2025} utilizes explainability mechanisms to explain \acp{CFG} for the purposes of malware detection. They utilize state-of-the-art techniques, such GNNExplainer and CaptumExplainer, in conjunction with a \ac{GNN} to generate a complex explanation of graphs used in the detection of malware. While this work is comprehensive and provides a detailed analysis of explainability techniques for graph-based \acp{GNN}, they do not extend the explainability mechanism to induce an adversarial attack on the trained model. Work from Aryal et al.~\cite{aryal2024explainability} utilizes explainability of model to craft adversarial attacks on byte-based malware detectors. Card et al.~\cite{card2024explainability} uses feature attribution to create adversarial attacks on dynamic feature based malware detectors. None of the work focuses on using XAI to graph-based malware detection systems.

The authors of \cite{soi_enhancing_2024} utilize graph-based information to augment explainability mechanisms in malware detection systems. By extracting features from the \ac{DEX} component of an \ac{APK} file, the authors are able to construct a graph that shows the relationship between system \ac{API} calls made by the application. Using this information, they're able to pair it with \ac{SHAP} to generate explanations of which functions contribute the most to a malicious classification. While the proposed system is useful for generating explanations for classifications, it fails to address the potential use for adversarial generation. Additionally, this framework is targeted at the Android \ac{OS} common to mobile devices instead of Windows.

\begin{figure}[!t]
    \centering
    \includegraphics[width=\columnwidth]{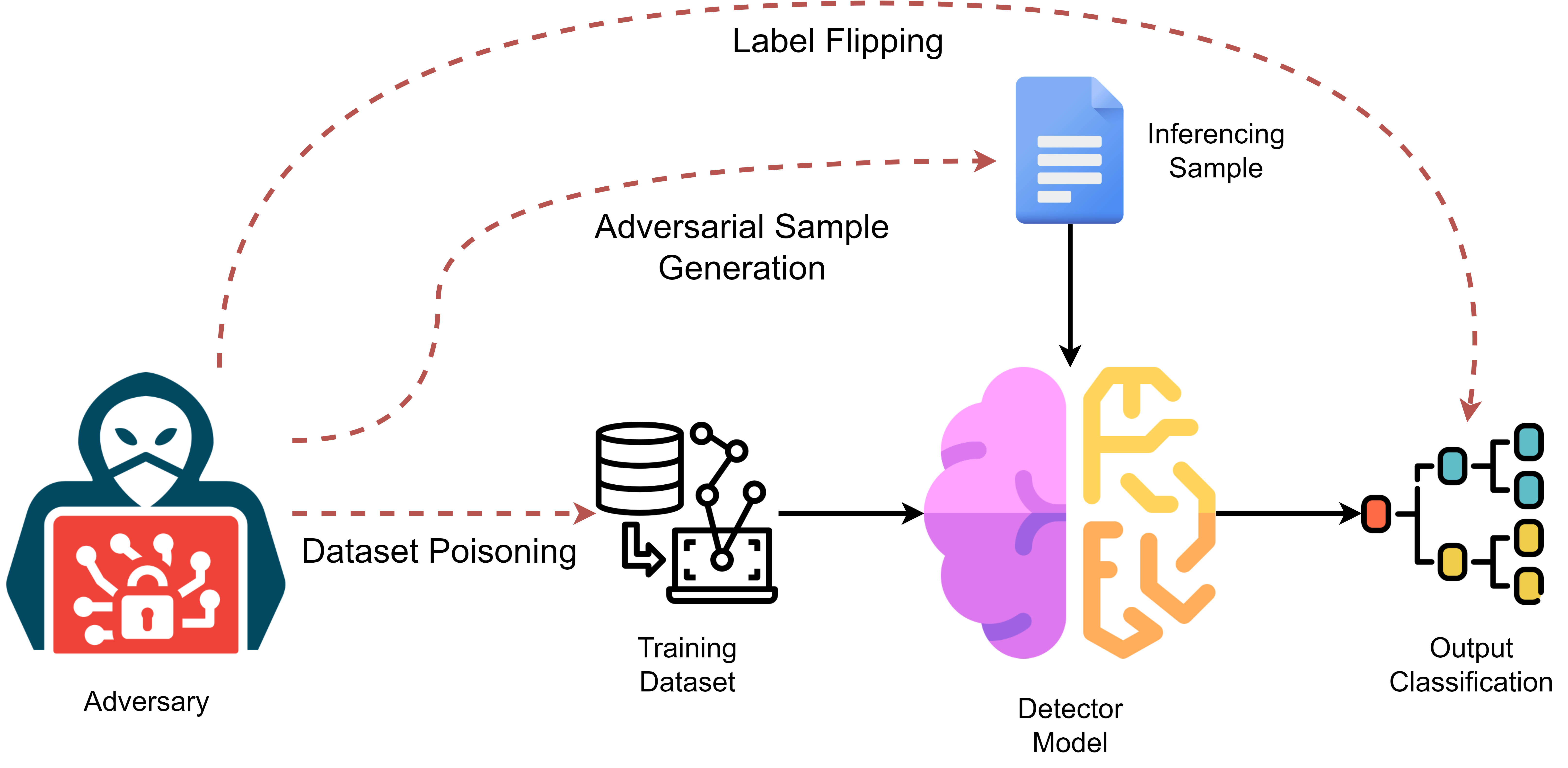}
    \caption{Generalized threat model of malware detection system.}
    \label{fig:ThreatModel}
\end{figure}

\subsection{Adversarial Attacks on Detector Systems}

In order to effectively deploy \ac{ML}-based detector systems in cybersecurity applications, the threat model needs to be analyzed. Figure \ref{fig:ThreatModel} describes the potential attack surface of \ac{ML}-based malware detection system. The attack surface includes all components that may be influenced by an adversary, including input data, training data, and the embedding layers of the model itself. Attacker capabilities range from black-box query access to full white-box knowledge of model parameters, directly shaping the sophistication of possible attacks. The level of access further distinguishes between remote interaction, local execution, or influence during training, which determines whether attacks such as evasion, extraction, or poisoning are feasible. Finally, practicality constrains adversarial strategies to those that preserve malware functionality and remain operationally realistic, ensuring that identified vulnerabilities reflect credible real-world risks rather than purely theoretical weaknesses.

In \cite{tagarelli_subgraph-based_2019}, the authors propose an adversarial attack framework designed to cause misclassifications in malware detection systems intended for \ac{IoT} systems. They accomplish this by modifying subgraph components of a program's \ac{CFG} in order to inject a small amount of perturbations into the sample. With this, they are able to create effective adversarial samples that require a minimal number of modifications. While this paper deals primarily with detector systems that utilize neural networks as the core architecture, it highlights the potential fragility in \ac{ML}-based malware detection systems.

The work of \cite{jiang_shield_2026} describes a black-box adversarial attack on \acp{LLM} designed to detect vulnerabilities in source code. By replacing sections of a source file with functionally-equivalent code, they're able to introduce vulnerabilities into a program that could be used by an attacker. This paper is significant as it specifically targets \acp{LLM} for the adversarial attack, but it focuses more on the introduction of vulnerabilities as opposed to the modification of already-existing executable files.

While there is a large corpus of existing work describing malware detectors using \ac{ML}, explainability of models trained on the \ac{CFG} modality, and adversarial attacks against detector systems, the intersection between all of these subjects is limited. Previous byte-level attacks \cite{kolosnjaji2018adversarial, aryal2024explainability} do not rely on a structural modality, and instead attempt to target the raw binary data present in an executable. Other approaches to graph-based model explainability \cite{yan_litexgnn_2025, shokouhinejad_consistency_2025} attempt no adversarial exploitation of the detector models. State-of-the-art attacks that target \acp{LLM} malware detectors \cite{jiang_shield_2026} often do so at the source code level, before the process that generates an Window \ac{PE} file.

\section{Proposed Methodology}\label{sec:Methodology}

\begin{figure*}[!t]
    \centering
    \includegraphics[width=0.75\textwidth]{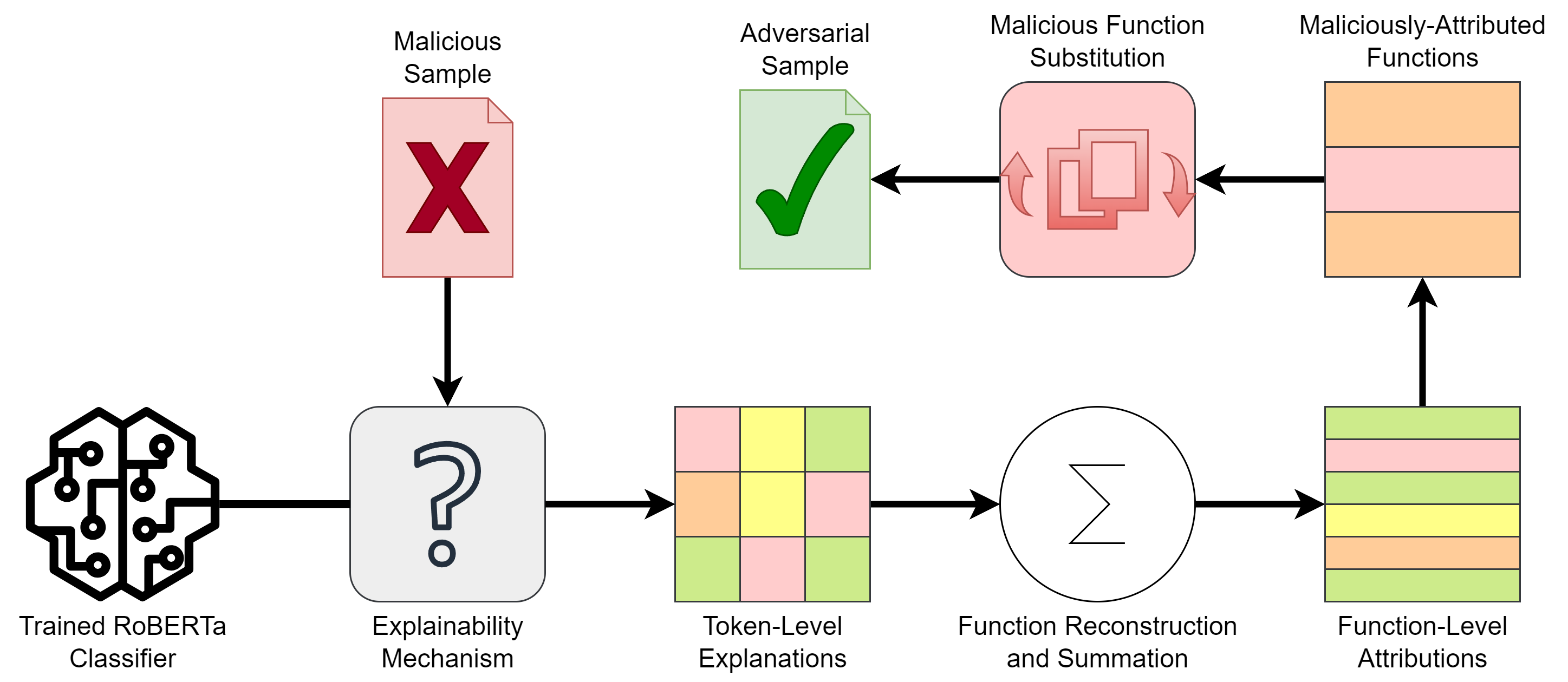}
    \caption{Overview of the adversarial generation procedure.}
    \label{fig:AttackArchitecture}
\end{figure*}

Figure \ref{fig:AttackArchitecture} details the general architecture for the proposed attack, and it highlights the components of the attack that occur after the model has been successfully trained. An explainability mechanism provided by the \emph{captum} \cite{captum} Python library generates sub-word token attributions. This library allows for easy integration to \emph{PyTorch}, which is the underlying library the \ac{RoBERTa} detector model is built on. The sub-word tokens can then be reconstructed into the original function name, and the constituent attributions can be summated to achieve a function-level attribtuion. We can then chose to modify the functions that contribute the most to a malicious classification via substitution to obtain an adversarial sample.


In order to ensure the model had a sufficient volume of data to train over, the dataset of \cite{lester_dataset_2021} was chosen. In addition to the dataset's size, the quantity of goodware files, which are often difficult to aggregate due to a wide variety of factors, is also highly desirable. Following dataset construction, the \ac{CFG} of each individual executable was extracted and converted into a \ac{JSON} file format before being written to disk. Graphs were traversed at runtime using a \ac{DFS} algorithm. 

\subsection{Dataset and Control Flow Graph}
During initial attack investigation, we experimented with different attack methodologies, such as different approaches to function name replacement, and their effects on model performance. To allow for more rapid iteration on the attack mechanisms and parameters, a smaller dataset was used. This dataset consists of 2,900 benign files and 6,679 malicious files. Benign files were sourced primarily from the system directories of a Windows 10 installation. Malware was sourced primarily from Malware Bazaar \cite{MalwareBazaar}. After this exploratory phase with the smaller dataset, a larger dataset \cite{lester_dataset_2021} was used to evaluate the model and the attack at scale. This dataset was then augmented by importing all samples from the small dataset, making the final consistency of the large dataset 94,123 malicious files and 54,154 benign files.

Once the dataset had been constructed and finalized, the \acp{CFG} were extracted using Python 3.11 and the \emph{r2pipe} \cite{radare2} module. In order to protect critical systems and mitigate any accidental infections, all feature extraction took place in a \ac{VM}. Resources from a host machine running Windows 11 were allocated using Oracle VirtualBox, and a total of four \ac{CPU} cores and 16 GiB of system memory were provisioned to allow for parallelized feature extraction. Debian 12.11 Bookworm was chosen as the guest \ac{OS}. Feature extraction took place at the function level, with all calls made to both internal subroutines and outside \ac{API} functions recorded in the graph data. The shallower analysis option \texttt{aa} for \emph{radare2} was used due to time and resource constraints. This level of analysis utilizes internal labeling information of the executable file to generate information, as opposed to the more advance \texttt{aaa} option, which utilizes complex algorithms for executable analysis. The graph information resulting from this operation was then saved to the disk in a \ac{JSON} format for future use. 


\subsection{RoBERTa Architecture}

\begin{figure*}[!t]
    \centering
    \includegraphics[width=0.8\textwidth]{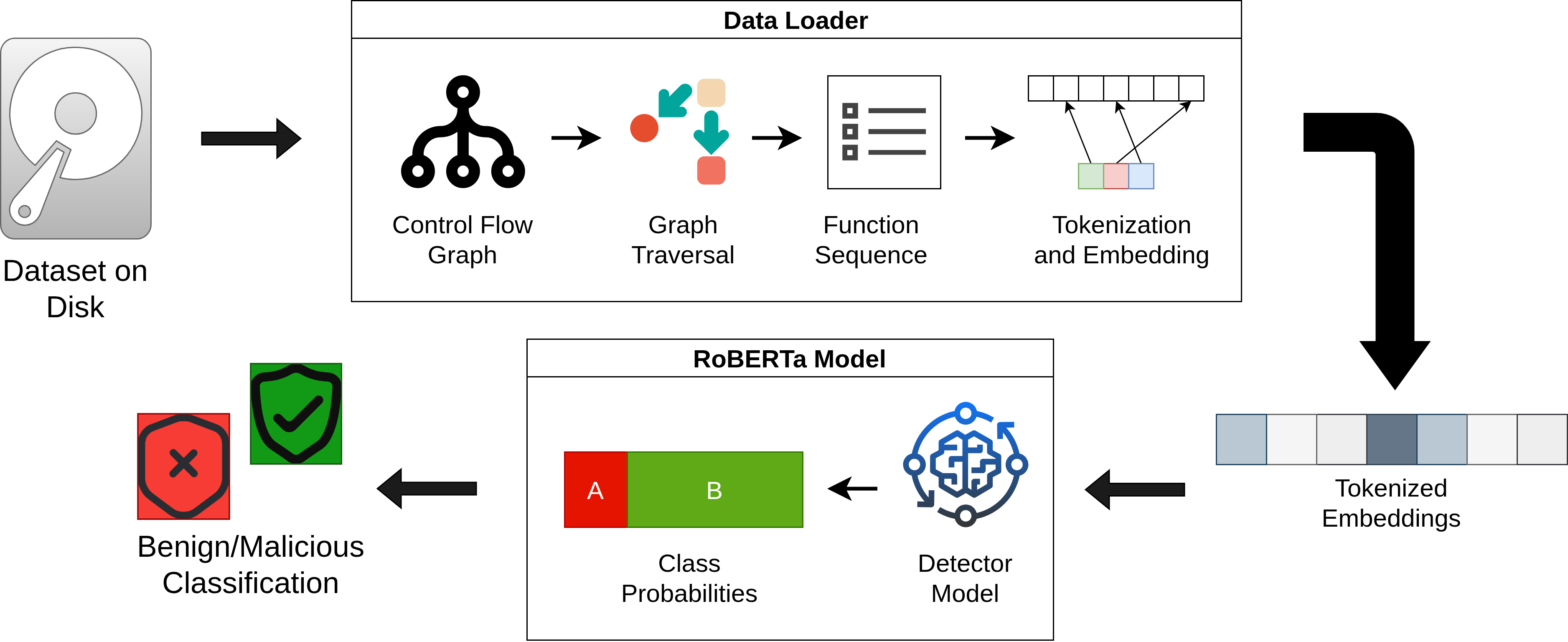}
    \caption{Architecture of the Graphene framework.}
    \label{fig:GrapheneArchitectureOverview}
\end{figure*}

Building from the work presented in \cite{wheeler_graphene_2026}, a pretrained \ac{RoBERTa} model can be adapted to the task of malware classification using \acp{CFG}. This reduces the amount of time needed to obtain a serviceable model by mitigating some development needs while also integrating a model that is proven to be effective for the task of malware classification into the pipeline. For this implementation, \ac{RoBERTa} can be thought of as two separate components: the \emph{data loader} and the \emph{model}. This is outlined in Figure \ref{fig:GrapheneArchitectureOverview}.

Graphene is a \ac{RoBERTa}-based malware detection frame that utilizes the control flow modality of Windows \ac{PE} files to detect if the corresponding file is benign or malicious. As can be seen in figure \ref{fig:GrapheneArchitectureOverview}, the \emph{data loader} ingests the executable from disk, converts it into a \ac{CFG}, obtains a linear function sequence, and embeds and tokenizes the functions. The model itself takes in the tokenized embeddings and generates a class probability between 0 and 1 for ``benign'' and ``malicious'' classes. These probabilities are then rendered as the final classification that indicates if the program is malware.

The data loader is the forward component of the model responsible for importing, parsing, and traversing the \ac{CFG} from a given input file. Data is loaded from \acp{CFG} already generated from executable files, meaning the graphs are already pre-extracted at the time of loading. The traversal step is needed to convert the nonlinear \acp{CFG} into a linear sequence of function calls stored as one large string. The framework of \cite{wheeler_graphene_2026} handles this by implementing a traversal handler in the data loader using the \emph{NetworkX} Python module. This handler automatically traverses the loaded graph and produces a string containing a series of function calls in the order they were encountered by the algorithm.

After a graph is converted into a string sequence by the data loader, the model converts the string into the form needed to train or inference. This is accomplished by using the \ac{RoBERTa} tokenizer that is paired with the pretrained model. Once all files in the dataset are loaded, the model is then trained further to differentiate between benign and malicious executable files utilizing their corresponding graph. The model was allowed to train for 5 epochs in all tests to ensure it had sufficient time to fit to the input data.

\subsection{Adversarial Sample Generation}
Generation of adversarial samples was performed in a white-box context. In this scenario, the adversary has full access to the target model, including its gradients, training parameters, and dataset. This means the attacker can generate an optimal adversarial sample without performing advance reconnaisance or training a proxy model to approximate the decision boundary of the target model. Selection of samples was random, where 2,500 total malicious samples were used as part of the attack. If a sample was benign, then it was not considered for selection.

Adversarial sample generation took place directly after model training completed. Like with a \ac{PGD} attack, the word replacement adversarial attack employed here utilized multiple rounds to iteratively generate the adversarial \ac{CFG}. This value was configurable at runtime, and it served as an upper limit for the number of attack iterations that could be deployed on a single sample. If a malicious sample was classified as benign before all attack rounds had been completed, all remaining rounds were skipped and the next sample was started. As is demonstrated in equation \ref{eq:RecursiveAdversarial}, if we take $A$ as the adversarial generation function, then it can be seen that the function is recursive in nature (\emph{i.e.} the output of the previous iteration serves as the input for the next).

\begin{equation}
    x_i =
    \begin{cases*}
        A(x_i) & if i = 0\\
        A(x_{i-1}) & if i $>$ 0
    \end{cases*}
    \label{eq:RecursiveAdversarial}
\end{equation}

Adversarial generation process is detailed in Algorithm \ref{alg:ExplainabilityAttack}. Gradients for the transformer were obtained using the \texttt{LayerIntegratedGradients} of the \emph{captum} \cite{captum} Python module. Token-level attribution was obtained using these gradients in \texttt{line 7}, and word-level attributions were obtained by calculating the sum of the word's token attributions in \texttt{line 8}. A positive attribution value for the word, and subsequently the corresponding function name, indicated that it contributed to the file being considered malicious, while a negative attribution was indicative of a benign classification. For every positively-attributed function, a replacement function in the form of a randomly-generated import was created, as is the case in \texttt{lines 12-13}. Each of these imports followed the general form of ``sym.imp.$x$.dll,'' where $x$ is a sequence of 10 random characters from a character pool of all capital letters and single-digit numbers. From the perspective of program structure, this is analogous to taking the particular function in question and moving its code to an external \ac{DLL} file. 

As can be seen in lines \texttt{15-17} in Algorithm \ref{alg:ExplainabilityAttack}, the resulting function sequence was provided to the target model again for inferencing, and the new classification was obtained. If this new classification was flipped to the benign classification, the attack was deemed a success and no further algorithm iterations were required. However, if the classification was still yielding the malicious class, and if the round counter hadn't reached the predefined limit, then the next round of the attack was performed on the already-modified sequence of function calls. Again, this process repeated until either the desired classification was obtained or no rounds were left to be performed. If the sample was still considered malicious at the end of all of the rounds, then the attack was considered a failure for that sample. Another potential case that could arise during the course of the attack is the summation of tokens into words could yield a set of words where none of them have a positive attribution. In these instances, the attack was stopped early as no progress could be made, and the attack was counted as a failure.

\begin{algorithm}
\caption{Explainability Attack}\label{alg:ExplainabilityAttack}
\begin{algorithmic}[1]
\State $failures = 0$
\State $successes = 0$
\State $attempts = 0$
\For {$sample \in dataset_{training}$}
    \If {$class(sample)$ \textbf{is} $malicious$}
        \For {$j \gets 0 < limit_{rounds}$}
            \State $attributions_{tokens} = explain(sample)$
            \State $attr_{words} = sum(attr_{tokens})$
            \If {$\forall a \in attributions_{words}, a \le 0$}
                \State $failures \gets failures + 1$
                \State $break$
            \Else 
            {$\forall a \in attributions_{words}, a > 0$}
                \State $word_a = replacement\_attack(word_a)$
            \EndIf
            \If {$class(adv\_sample_i)$ \textbf{is} $benign$}
                \State $successes \gets successes + 1$
                \State $break$
            \EndIf
        \EndFor
        \State $attempts \gets attempts + 1$
        \If{$attempts = limit_{rounds}$}
            \State $stop$
        \EndIf
    \EndIf
\EndFor
\end{algorithmic}
\end{algorithm}

\section{Experimental Results and Analysis}\label{sec:Results}

For training the malware detection model, the datasets were split in a 4:1 ratio with 80\% of the samples being used for actual training and the remainder for evaluation. The evaluation process was performed at the end of each epoch to gauge model performance as it trained.

\subsection{Experimental Setup}

Training of the \ac{RoBERTa} model was performed on a \ac{HPC} cluster with 8 \acp{CPU} cores and 64 GiB of system memory. For training on the smaller dataset, only a single \ac{GPU} was allocated, but for the larger dataset, two \acp{GPU} were allocated and the model weights synchronized across each unit at the end of each epoch. The second \ac{GPU} was needed due to the large amount of data to be processed in the larger dataset.

Following in the work of \cite{wheeler_graphene_2026}, training parameters were set to similar values to ensure repeatable experimentation results. A batch size of 64 was used due to \ac{GPU} memory and processing constraints, and a total of 16 function calls from each \ac{CFG} were used. Because the previous work found relatively negligible differences in model performance between traversal algorithms, the \ac{DFS} algorithm was used to generate the linear sequence of function calls from the non-linear graph. The model was trained for 5 epochs to ensure it had a sufficient amount of time to adjust to the dataset. During training, models had an accuracy of 95.35\% for the small dataset and 94.69\% for the large dataset, with a \ac{FPR} or 3.57\% and 3.63\%, respectively.

When launching the attacks against the trained \ac{RoBERTa} models, several metrics were collected. For attack failures, two separate metrics were tracked. Firstly, we tracked the total number of failures across all samples that were partitioned from the dataset for the attack, and secondly, we excluded all samples that were considered unimprovable from this set. That is, samples who had no functions that were positively attributed to a malicious classification were excluded from the second count. A distinction between these two methods of metric collection was made as one captures what would likely be real-world attack success while the other records the success rate of the attack in a vacuum without samples that are incompatible with the algorithm. For the attribution of tokens using the \texttt{LayerIntegratedGradients} component of \emph{captum} the default value of 50 was used, as it provides a strong balance between computational overhead and approximation accuracy.

\subsection{Attack Performance}
The general success rate of the attack $s_g$ was calculated by dividing the total number of successful attacks $a_{s}$ by the total number of attacks attempted $a_{a}$, as can be seen in equation \ref{eq:GeneralSuccess}. A second success rate, the success rate of attacks excluding samples that could not be improved $s_n$, was calculated by dividing the total number of successful attacks $a_{s}$ by the total number of attacks attempted that did not involve a sample that could not be improved $a_{i}$, as is demonstrated in equation \ref{eq:ImprovableSuccess}. By necessity, $a_{a} \ge a_{i}$, as there is always the possibility that some samples cannot be improved.

\begin{equation}
    s_g=\frac{a_{s}}{a_{a}}
    \label{eq:GeneralSuccess}
\end{equation}

\begin{equation}
    s_n=\frac{a_{s}}{a_{i}}
    \label{eq:ImprovableSuccess}
\end{equation}

Attack iterations were varied between 1, 2, 3, and 5, allowing for a wide array of attack scenarios to be tested. For each iteration value, a total of 3 trials were conducted to obtain a grounded baseline for attack success patterns. During each attack, metrics for $s_g$ and $s_n$ were collected, along with the total number of improvable samples. Each set of iteration values was conducted on the small and large datasets.

For a RoBERTa model that trained on the small dataset, the attack was generally successful, both in regards to $s_g$ and also $s_n$. Additionally, the mean and median for all attacks was generally within a few percentage points of each other, as can be seen in Table \ref{tab:RoBERTaMetricsSmall}. The value for $a_{i}$ was never below 2,200, meaning that roughly 90\% of the time the attack was able to attempt to generate an adversarial sample at least once. 

The worst-performing attack for the small dataset was the single-iteration attack with an average $s_g$ of 87.72\%. This is to be expected because the likelihood that the attack can generate a usable adversarial sample is lower the fewer rounds are available to the attack. However, if we exclude all unimprovable samples, the success rate $s_n$ jumps to an average of 97.24\%. The lowest average $s_n$ was for the two-round attack, at 93.81\%. Because this value deviates from the median by a relatively significant margin when compared to the other attacks, this lowered average is most likely due to an outlier being present in the collected data.

When the model was trained using the larger, attack statistics dropped dramatically, as can be seen in Table \ref{tab:RoBERTaMetricsLarge}. The best performing attack was using 3 iterations, with an average general success rate of 69.88\%. Again, excluding unimprovable samples, we see the success rate increase to as high as 97\%. The lowest average $s_n$ for the attack on the larger dataset was the single iteration attack with an average of 85.40\%.

Malicious samples that were considered unimprovable most likely had sub-function name tokens that contributed greatly to their malicious classification, but when those tokens were aggregated back into function names, the summation indicated the function was benign.

\subsection{Results Discussion and Analysis}

\subsubsection{Attack Efficacy}
\begin{table}[]
\centering
\begin{tabular}{c|cccccc}
Rounds & \multicolumn{2}{c|}{$s_g$}                                  & \multicolumn{2}{c|}{$a_i$}                              & \multicolumn{2}{c}{$s_n$}              \\ \hline
       & \multicolumn{1}{c|}{Mean}    & \multicolumn{1}{c|}{Median}  & \multicolumn{1}{c|}{Mean} & \multicolumn{1}{c|}{Median} & \multicolumn{1}{c|}{Mean}    & Median  \\ \hline
1      & \multicolumn{1}{c|}{87.72\%} & \multicolumn{1}{c|}{87.88\%} & \multicolumn{1}{c|}{2255} & \multicolumn{1}{c|}{2249}   & \multicolumn{1}{c|}{97.24\%} & 96.61\% \\
2      & \multicolumn{1}{c|}{88.52\%} & \multicolumn{1}{c|}{90.40\%} & \multicolumn{1}{c|}{2365} & \multicolumn{1}{c|}{2318}   & \multicolumn{1}{c|}{93.81\%} & 97.50\% \\
3      & \multicolumn{1}{c|}{91.44\%} & \multicolumn{1}{c|}{90.04\%} & \multicolumn{1}{c|}{2300} & \multicolumn{1}{c|}{2285}   & \multicolumn{1}{c|}{99.37\%} & 99.77\% \\
5      & \multicolumn{1}{c|}{90.81\%} & \multicolumn{1}{c|}{90.60\%} & \multicolumn{1}{c|}{2278} & \multicolumn{1}{c|}{2271}   & \multicolumn{1}{c|}{99.65\%} & 99.74\%
\vspace{1mm}
\end{tabular}
\caption{Attack statistics for RoBERTa trained on small dataset.}
\label{tab:RoBERTaMetricsSmall}
\end{table}

Attack performance when the model was allowed to train on a much large dataset (Table \ref{tab:RoBERTaMetricsLarge}) indicates that the attack is still feasible, albeit with reduced efficacy. Again, the worst-performing attack was the one that used a single iteration. Average $s_g$ for it was 61.51\%, with some attacks have a general success rate as low as 53\%. Attack performance did improve as the number of iterations increased, but this improvement was limited and it even dropped between 3 and 5 iterations. This shows the attack potential could be limited, and the number of iterations may have a soft cap for improvement before the changes made are more of a detriment to the adversary.

\begin{table}[]
\centering
\begin{tabular}{c|cccccc}
Rounds & \multicolumn{2}{c|}{$s_g$}                                  & \multicolumn{2}{c|}{$a_i$}                              & \multicolumn{2}{c}{$s_n$}              \\ \hline
       & \multicolumn{1}{c|}{Mean}    & \multicolumn{1}{c|}{Median}  & \multicolumn{1}{c|}{Mean} & \multicolumn{1}{c|}{Median} & \multicolumn{1}{c|}{Mean}    & Median  \\ \hline
1      & \multicolumn{1}{c|}{61.51\%} & \multicolumn{1}{c|}{61.20\%} & \multicolumn{1}{c|}{1802} & \multicolumn{1}{c|}{1867}   & \multicolumn{1}{c|}{85.40\%} & 86.86\% \\
2      & \multicolumn{1}{c|}{65.53\%} & \multicolumn{1}{c|}{64.40\%} & \multicolumn{1}{c|}{1720} & \multicolumn{1}{c|}{1733}   & \multicolumn{1}{c|}{95.34\%} & 95.80\% \\
3      & \multicolumn{1}{c|}{69.88\%} & \multicolumn{1}{c|}{69.76\%} & \multicolumn{1}{c|}{1804} & \multicolumn{1}{c|}{1798}   & \multicolumn{1}{c|}{96.95\%} & 97.00\% \\
5      & \multicolumn{1}{c|}{67.36\%} & \multicolumn{1}{c|}{70.96\%} & \multicolumn{1}{c|}{1911} & \multicolumn{1}{c|}{2031}   & \multicolumn{1}{c|}{88.21\%} & 88.23\%
\vspace{1mm}
\end{tabular}
\caption{Attack statistics for RoBERTa trained on large dataset.}
\label{tab:RoBERTaMetricsLarge}
\end{table}

However, even if the attack potential is reduced, the attack still has applicability as a potential threat to the detector model. The average value of $a_i$ for the large dataset indicates that there is still at least an approximately 70\% chance that the sample submitted can have the attack employed on it, and of that, there is an 85\%-96\% chance that the adversarial sample generated can cause the generated sample to be misclassified. For an advance and robust detector model such as \ac{RoBERTa}, being able to generate an adversarial sample so consistently is difficult in and of it self.

Which functions are considered to be malicious is often dependent on the distribution of the function names as they are related to the goodware and malware. The more often a function is called, the higher the likelihood that it is called by a malicious file. Table \ref{tab:CallsSmallDataset} provides the top ten most common function calls made within the small dataset. 


\begin{table}[]
\centering
\begin{tabular}{c|c}
\hline
Function Call                          & Function Count  \\ \hline
entry0                                 & 5307            \\
sym.imp.KERNEL32.dll\_GetStartupInfoA  & 3153            \\
sym.imp.KERNEL32.dll\_GetModuleHandleA & 2631            \\
sym.imp.MSVCRT.dll\_\_set\_app\_type   & 2282            \\
sym.imp.MSVCRT.dll\_\_\_p\_\_fmode     & 2282            \\
sym.imp.MSVCRT.dll\_\_\_p\_\_commode   & 2282            \\
sym.imp.MSVCRT.dll\_\_\_setusermatherr & 2282            \\
sym.imp.MSVCRT.dll\_\_\_getmainargs    & 2279            \\
sym.imp.MSVCRT.dll\_exit               & 2273            \\
unk.0x40a411                           & 1781
\vspace{1mm}
\end{tabular}
\caption{Top ten most common function calls in small dataset.}
\label{tab:CallsSmallDataset}
\end{table}

In Table \ref{tab:CallsSmallDataset}, the most common function name that appears is \emph{entry0}. This is not an anomalous result as this function often serves as the entry point to the executable code itself. The \emph{GetStartupInfoA} and \emph{GetModuleHandleA} are utility functions commonly called to obtain environment parameters, which can influence how an executable should behave. Again, these functions are common to both benign and malicious file alike. The functions originating from \emph{MSVCRT.dll} are part of Microsoft's Visual C Runtime and are used in the capacity of utility functions as well.

The final entry in the table, \emph{unk.0x40a411}, is an unusual occurrence. This function name indicates that a subroutine that is not external to the executable exists at address 0x40a411 within it's code. Most likely, this comes from the fact that there are structural similarities shared between a large number of executable files. These similarities could arise from a number of origins, such as compiler-generated code consistently aligning functions to that address. If the majority of samples that fall into this category are malware, this repeating address could be from the samples all belonging to the same family of malware, or derivatives of similarly-behaving programs. More detailed analysis is needed to accurately identify the cause of this occurrence.

\subsubsection{Attack Practicality}
While converting an adversarial sequence of function calls back into an executable program may be difficult to achieve, it is theoretically possible given the correct toolkit. In order to create a sample that has the same resulting control flow graph as the adversarial sequence of function calls, the components in code that correspond to the internal functions would need to be moved to an external \ac{DLL} file. If the original source code of the program is present, this is a relatively simple process of writing an independent program that can be compiled into a \ac{DLL}, adding the code in question, and importing it at runtime using the Windows \ac{API}. The practicality of this endeavor may become more difficult if a large number of functions have to be moved to external files, which could lead to broken dependencies. 



\section{Conclusion}\label{sec:Conclusion}
This work investigated the applicability and efficacy of a white-box adversarial evasion attack using explainability mechanisms for a \ac{RoBERTa}-based malware detection model. Using a model that has proven to be resilient and robust in its detection capabilities, the corresponding modality, a \ac{CFG}, was modified via a multi-iteration attack, and the model's ability to determine if the sample was benign or malicious was tested. We determined that exporting internal function calls to external \acp{DLL} was an effective way of decreasing detection rate for the \ac{RoBERTa}-based system. We reached a high attack success rate close to 100\% for small dataset and 94\% for larger dataset.


Looking to future works, expanding the \ac{RoBERTa} detector system to include both Windows and Linux executable files could help determine if this modality is sufficient enough to abstract away from \ac{OS}-specific modalities. The attack could then be extended to handle call graphs for programs that don't use \ac{DLL} files as imports.

\section*{Acknowledgement}
The authors acknowledge the Research Computing and Data division at Tennessee Tech University (RRID:SCR\_027555) for providing computational resources and support services that have contributed to the research results reported within this paper. This work is partially supported by the NSF grants 2230609, 2416990 and 2346001, and DoD CSA HQ00342410054 at Tennessee Tech University.

\bibliographystyle{./IEEEtran}
\bibliography{./References.bib}

\end{document}